\documentclass{ws-ijmpe}

\def\nuc#1#2{$^{#1}$#2}

\begin{document}

\markboth{L. Pr\'ochniak}{%
Collective excitations of transactinide nuclei 
in a self-consistent mean field
theory}

\catchline{}{}{}{}{}

\title{\MakeUppercase{%
Collective excitations of transactinide nuclei in a self-consistent mean field
theory
}
}

\author{\footnotesize \MakeUppercase{L. Pr\'ochniak}}

\address{Institute of Physics, Maria Curie-Sk{\l}odowska University,
 pl. M.Curie-Sk{\l}odowskiej~1,\\ 20--031 Lublin, Poland}

\maketitle

\begin{history}
\received{(October 31, 2007)}
\revised{(revised date)}
%\accepted{(Day Month Year)}
%\comby{(xxxxxxxxxx)}
\end{history}

\begin{abstract}
We applied the ATDHFB approach for  study of properties of collective
quadrupole states in several transactinide nuclei: \nuc{238}{U}, \nuc{240}{Pu}, \nuc{242}{Pu},
\nuc{246}{Cm}, \nuc{248}{Cm}, \nuc{250}{Cf} and \nuc{252}{Cf}. 
Calculated  energies and B(E2) transition probabilities are in a reasonable
agreement with experimental data. We present also  results concerning
superdeformed collective states in the second minimum of potential energy
of the \nuc{240}{Pu}
nucleus.
\end{abstract}

\section{Introduction}

Fission properties of heavy and superheavy nuclei have been recently a
subject of intensive studies within the frame of a self-consistent mean 
field theory. Most papers treated height of fission
barriers\cite{2002WA22,2004BO34,2004BU03,2006ST04,2006DE23} but some results
on half-lives of superheavy nuclei have also been published.\cite{2007BA17}
The standard method of calculating the half-lives is the WKB(J) approach
requiring knowledge of the potential and mass parameter(s), which are
usually obtained from the Adiabatic Time Dependent Hartree-Fock-Bogolyubov
(ATDHFB) or the Generating Coordinate Method (GCM) theory. These general
theories can provide a frame for a description of other collective
phenomena, e.g. quadrupole rotational-vibrational excitations. However such
applications of ATDHFB or GCM in the region of transactinides are rather
scarce.\cite{2006DE23} On the other hand collective levels and E2 transition
probabilities of lighter transactinides (U,Pu,Cm) are well known
experimentally. Moreover thanks to recent advances in experimental
techniques the region of heavier nuclei accessible for measurements is
growing rapidly.  In this paper we present several results of application of
the ATDHFB theory with the Skyrme interaction for describing quadrupole
collective properties of \nuc{238}{U}, \nuc{240}{Pu}, \nuc{242}{Pu},
\nuc{246}{Cm}, \nuc{248}{Cm}, \nuc{250}{Cf} and \nuc{252}{Cf}
 nuclei, for which there is an abundant  experimental data.
The main part of the paper is devoted to normally deformed states, i.e., built
around the first minimum of the potential energy but
Subsection~\ref{subs:sd} contains also  results of calculation of
superdeformed (connected with the second minimum of the potential) states in
the \nuc{242}{Pu} nucleus.

\section{Theory}
The main tool used in microscopic theory of full five-dimensional quadrupole
collective dynamics is the ATDHFB theory, which leads to the generalized Bohr
Hamiltonian when appropriate collective variables are chosen. The collective
Hamiltonian contains potential energy and mass parameters (including moments
of inertia) calculated solely from a microscopic input, i.e., nucleon-nucleon
interaction. Details of the method we use and results obtained for lighter
nuclei can be found in.\cite{2004PR01,2007PR11} Below we recall only some of main
points.   

Collective quadrupole variables
$\beta$, $\gamma$
are defined by components of the second rank tensor of a nuclear mass
distribution:
\begin{equation}
\begin{array}{ll}
\beta\cos\gamma=q_0\sqrt{\pi/5}/A\langle r^2\rangle & 
\langle r^2\rangle=\frac{3}{5} (r_0 A^{1/3})^2, \ \ r_0=1.2 {\rm fm}\\ 
\beta\sin\gamma=q_2\sqrt{3\pi/5}/A\langle r^2\rangle\\
q_0=\langle Q_0 \rangle=\langle \textstyle\sum_i 3z^2_i-r_i^2 \rangle\\
q_2=\langle Q_2 \rangle =\langle\textstyle \sum_i x^2_i-y_i^2 \rangle \ .
\end{array} 
\end{equation}
Hence in the axial case $\beta$ for a given nucleus is strictly proportional
to the mass quadrupole moment. The self-consistent nuclear mean field is
obtained from HF+BCS calculations with a double constraint: 
\begin{equation}
\delta\langle H - \lambda_0 Q_0 -\lambda_2 Q_2\rangle=0 \ .
\end{equation}
Then the ATDHFB theory allow  to calculate mass parameters and collective
potential energy that enter the Bohr Hamiltonian. Its eigenvalues are
interpreted as collective excitation energies and its eigenfunctions
can be used, among others, to obtain E2 transition probabilities.

\subsection{Details of calculations}

We have chosen as a microscopic interaction the SkM* version of Skyrme
forces, proposed already long time ago\cite{1982BA39} but still regarded as
a good choice especially in the case of barrier heights and other fission
properties.\cite{2004BO34,2006ST04,2007BA17} We have performed also
calculations with the SIII Skyrme interaction which previously gave good
results for lighter nuclei\cite{2004PR01,2007PR11} and a few tests with the
more recent SLy4 forces.  In the particle-particle channel we have taken the
simplest form of the pairing interaction i.e. of seniority type (constant
$G$). Strength of the pairing was fixed by comparing values of the 5-point
formula pairing gap with minimal quasiparticle energies at points
corresponding to a minimum of the potential in U and Pu nuclei. Final
results for the strength are $G_{\rm n,p}=g_{\rm n,p}/(11+N(Z))$, $g_{\rm
n}=15.1$~MeV, $g_{\rm p}=14.9$~MeV. As we checked in some test cases the
results from a state dependent ($\delta$) pairing interaction are almost the
same as from the constant $G$ force, provided the strength of the
interaction is fixed using the same method.

The calculations were made for  220 points in the sextant ($0\le \gamma \le
60^{\circ}$, $0\le \beta \le 1$) of the deformation plane. These point form
a regular grid with the distance of 0.05 and $6^{\circ}$ between them. The
maximal value of $\beta$ (i.e. $\beta=1$) corresponds for the considered
nuclei to a quadrupole moment around 100~b. For the lowest normally deformed
collective states only the region $\beta \lesssim 0.6$ is important but for
studying the second minimum one must expand the range of $\beta$
considerably.

It must be stressed that we exclude the octupole deformation which is
essential for fission processes, but for the considered nuclei it becomes
important only for larger deformations than those studied by us. As we
discussed previously in\cite{1999PR03,2004PR01,2006PR05}, see
also\cite{1999LI38}, pairing vibrations and so called Thouless-Valatin
corrections can have a noticeable influence on the mass parameters. In the
present work we took a simplified approach introducing an average factor 1.3
by which we multiply values of the mass parameters obtained from the ATDHFB
formulas.

\begin{figure}[htb]
\centerline{\includegraphics[scale=0.45]{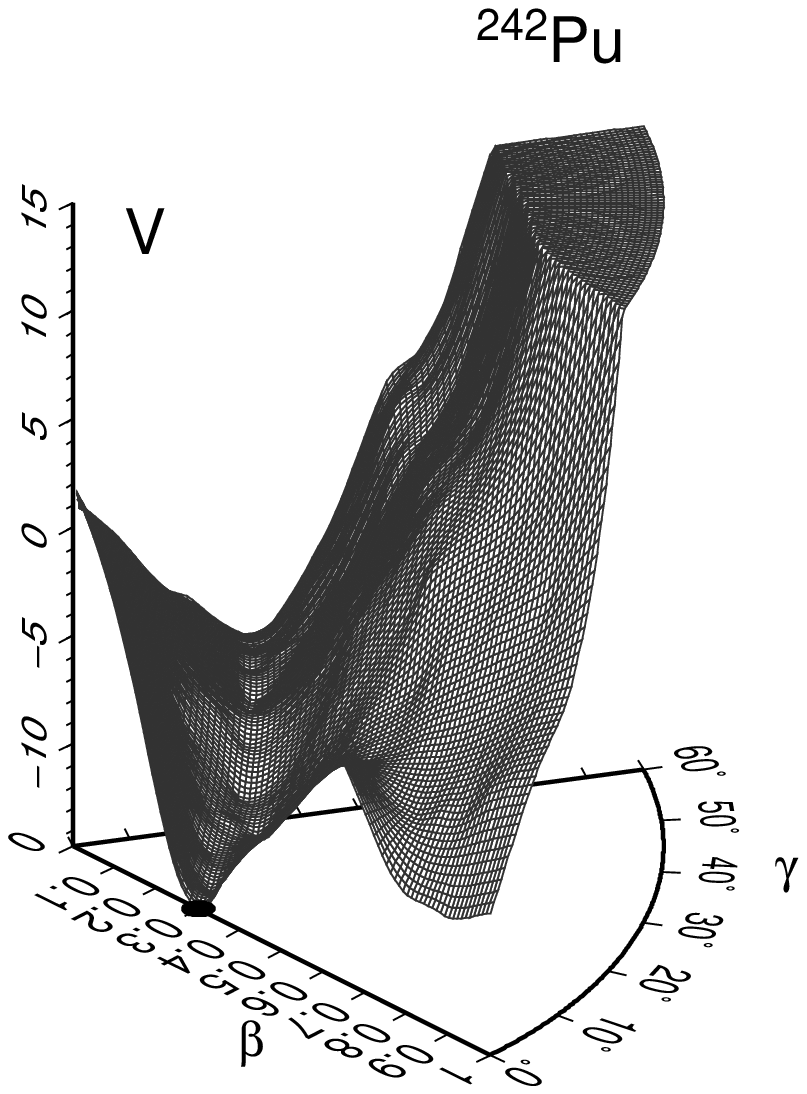}\hspace{2cm}
\includegraphics[scale=0.45]{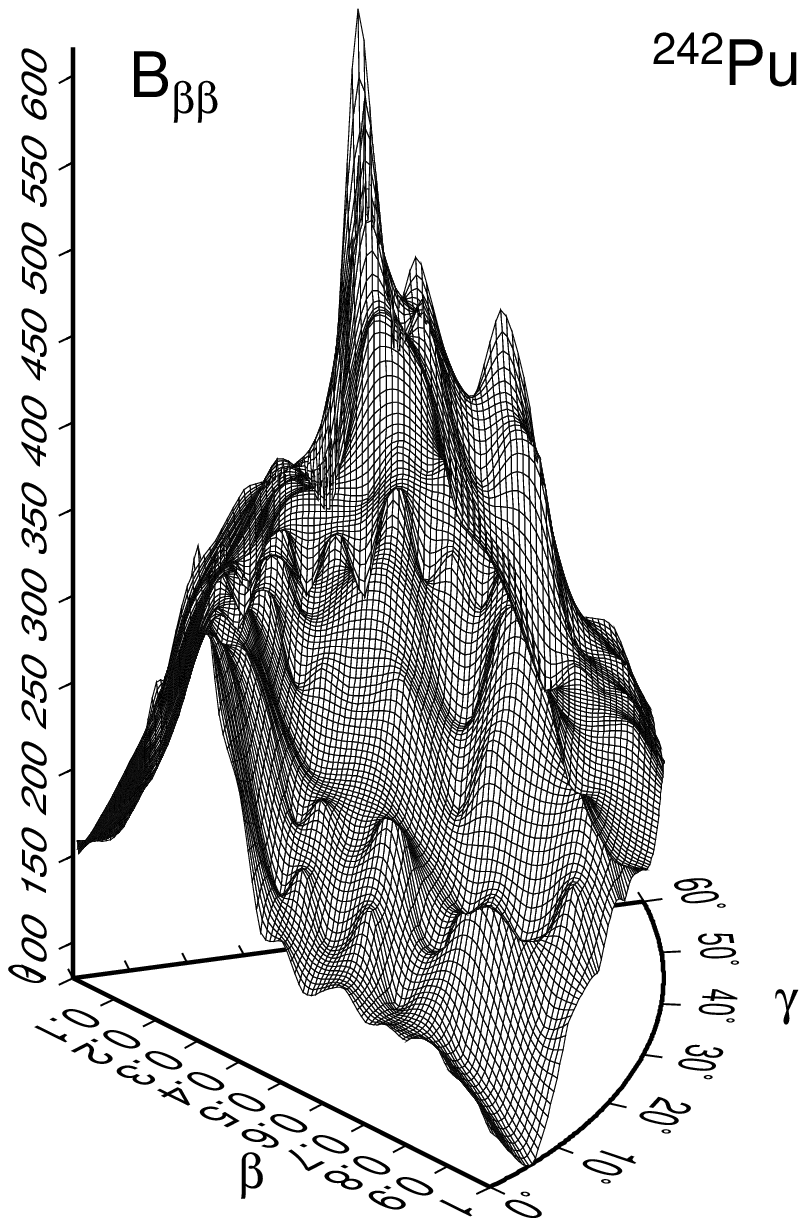}}
\caption{Collective potential energy $V$  (left panel) and the  mass parameter
$B_{\beta\beta}$ (right panel) for the \nuc{242}{Pu} nuclei.\label{fig:potbbb}}
\end{figure}

For
each nuclei we calculate seven functions (the potential energy  and six mass
parameters, including moments of inertia). In~Fig.~\ref{fig:potbbb} 
we show only a small sample, namely  the potential energy  $V$ and the mass
parameter $B_{\beta\beta}$ for the \nuc{242}{Pu} nucleus.  

\section{Results}

\subsection{Energy levels}

\begin{figure}[htb]
\centerline{\includegraphics[scale=0.35]{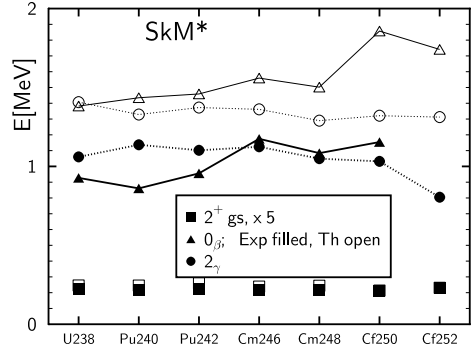}} \caption{Selected
experimental{\protect\cite{nndc0907}} (filled symbols) and theoretical (open
symbols) energy levels ($2^+_{\rm gs}$, $\beta$ and $\gamma$ bandheads). The
energy of the $2^+_{\rm gs}$ level is multiplied by 5.\label{fig:levskm}}
\end{figure}

In Fig.~\ref{fig:levskm} we show a calculated energy of the first $2^+_{\rm
g.s.}$ level and of bandheads of $\beta$ and $\gamma$ bands and we compare
them with experimental data.\cite{nndc0907} Please note that the energies
(theoretical and experimental) of the $2^{+}_{\rm gs}$ level are multiplied
by 5 to make the figure more readable. 
Keeping in mind the starting point of calculations which was
the nucleon-nucleon effective interaction and the fact that we did not fit
any parameter to analyzed data, the results shown in Fig.~\ref{fig:levskm}
are quite good, even despite too large energies of the theoretical $\beta$
and $\gamma$ bandheads.

\begin{figure}[htb]
\centerline{\includegraphics[scale=0.35]{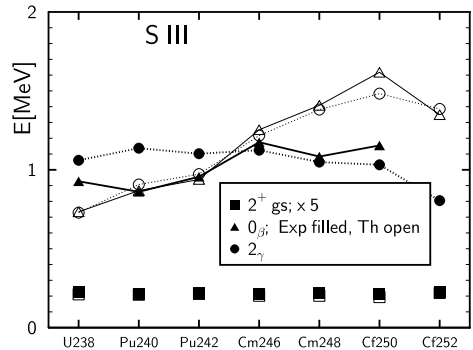}}
\caption{See caption to Fig.~\ref{fig:levs3}, theoretical levels are obtained
using the SIII Skyrme interaction.\label{fig:levs3}}
\end{figure}

\begin{figure}[htb]
\centerline{\includegraphics[scale=0.35]{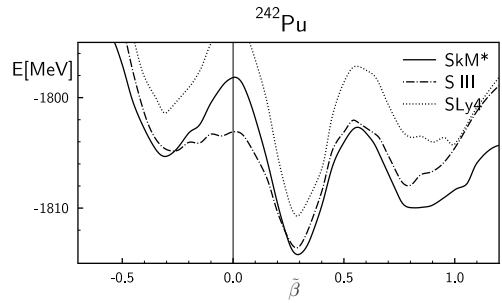}}
\caption{Comparison of the potential energy calculated for axial shapes of the
\nuc{242}{Pu} nucleus using various Skyrme interaction. In this figure
negative values of $\tilde{\beta}$ correspond to an oblate shape.\label{fig:porpot}}
\end{figure}

Below we present also analogous results  for the SIII Skyrme interaction
(Fig.~\ref{fig:levs3}). As it can be seen the $\beta$ and $\gamma$ bandheads
are now on average closer to experimental values, 
but their behavior with an increasing
mass number does not follow the experimental one.  
Detailed inspection of calculated potential energy and mass parameters shows
that differences between the results of the SkM* and SIII variant of Skyrme interaction 
stem mainly from different behavior of the potential in the region of small
deformations. It can be readily seen in Fig.~\ref{fig:porpot}, where we plot
the potential energy for axial shapes of the \nuc{242}{Pu} nucleus obtained
using the SkM*, SIII and SLy4 forces. This figure shows also that dependence of
$V$ on deformation for the SkM* and SLy4 forces is very similar.

Let us mention another remarkable point. In many papers the energy of the
first $2^+$ level is estimated from the formula $E_{2,\rm rot}=2(2+1)/2J$,
where $J$ is the moment of inertia calculated in the minimum of the
potential, in other words assuming a perfect rotor behavior. Within the
frame of our approach we can compare the value obtained in such a way with
the respective eigenvalue of the Bohr Hamiltonian $E_{2, \rm Bohr}$. It
appears that $E_{2, \rm Bohr}$ is grater typically by 16--20~keV than
$E_{2,\rm rot}$. This correction comes from vibrational degrees of freedom
and is very small if compared with the energy of, loosely speaking, $\beta$
and $\gamma$ phonons but is quite large in relation to $E_{2, \rm exp}$.

\subsection{E2 electromagnetic transitions}

\begin{figure}[htb]
\centerline{\includegraphics[scale=0.35]{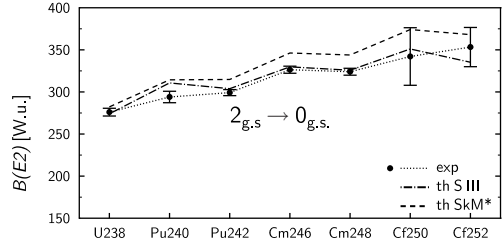}}
\caption{$B(E2)$ probabilities for transition 
$2_{\rm gs}\rightarrow 0_{\rm gs}$. 
Experimental data taken from{\protect\cite{nndc0907}}.\label{fig:e2gs}}
\end{figure}

\begin{figure}[htb]
\centerline{\psfig{file=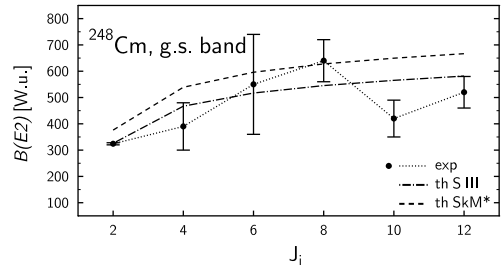,scale=0.35}}
\caption{$B(E2)$ transition probabilities within the ground state band in
the \nuc{248}{Cm} nucleus. $J_i$ denotes spin of an initial level.\label{fig:e2cm248}}
\end{figure}

In this subsection we compare theoretical $B(E2)$ transition probabilities
with the experimental ones, taken from.\cite{nndc0907} Fig.~\ref{fig:e2gs}
shows transitions $2_{\rm gs}\rightarrow 0_{\rm gs}$ in the considered
nuclei while Fig.~\ref{fig:e2cm248} contains results concerning transitions
within the ground state band in the \nuc{248}{Cm} nucleus. Moreover in
Table~\ref{tab:e2_inter} we present some inter-band transitions in the
\nuc{250}{Cf} nucleus.  

\begin{table}[htb]
\tbl{$B(E2)$ probabilities (W.u.) for inter-band ($\gamma \rightarrow $ {g.s.}) 
transitions in $^{250}$Cf.\label{tab:e2_inter}}
{\begin{tabular}{lllrr}\toprule
&&              Exp     &             SkM* & S III\\
\colrule
$2^+$ ($\gamma$ band) $\rightarrow$ g.s. &$4^+$&0.21 $\pm$ 0.02&  0.87 &
0.49\\
			&$2^+$&     3.7  $\pm$ 0.4&               10.32 &
8.16\\
			&$0^+$&     2.3  $\pm$ 0.3&               6.26 &
4.73\\
\botrule
\end{tabular}
}
\end{table}
 
The conclusion is that theoretical results follow experiment very closely.
We recall again that we do not use here any additional parameters (effective
charge etc.). And again the SIII interaction performs slightly better than
SkM*.

\subsection{Superdeformed collective states\label{subs:sd}}

Theory presented in the previous sections gives also possibility to study
superdeformed collective states i.e. built in the vicinity of the second
minimum of the potential energy. It needs however a large extension of the
basis used to solve the eigenproblem of the Bohr Hamiltonian. We presented
such calculations in the approach with the phenomenological Nilsson
potential in Ref.~\cite{2002PR01}, see also papers by Libert {\it et al.}
using Gogny forces.\cite{1999LI38,2006DE23} In the present paper we show
results for the \nuc{242}{Pu} nucleus, for which there are many experimental
data\cite{2000PA40,2001GA05,2001HU12}, see also review in
Ref.~\cite{2006DE23} Moreover in this nucleus an outer barrier is
sufficiently high (see also Fig.~\ref{fig:porpot}) so as we do not need to
worry about e.g. coupling with the continuum.
   
\begin{figure}[htb]
\centerline{\includegraphics[scale=0.45]{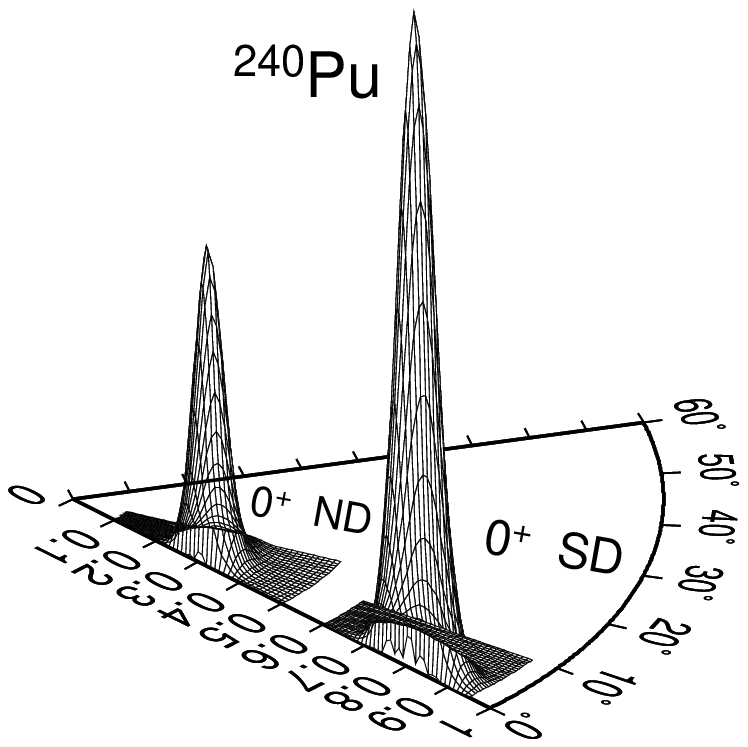}\hspace{2cm}
\includegraphics[scale=0.45]{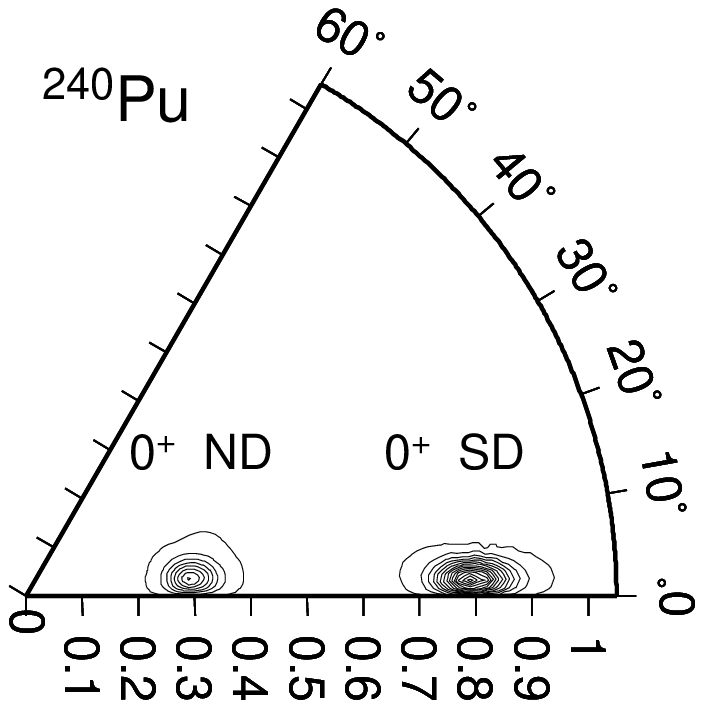}}
\caption{Probability distribution $|\Psi|^2\det g$ for the lowest $0^+$
normal (i.e. ground) and superdeformed state.\label{fig:funfal}}
\end{figure}

Fig.~\ref{fig:funfal} shows the probability distribution for the ground
state and the lowest $J=0$ state that can be unambiguously identified as a
superdeformed one. More precisely we plotted here the product
$|\Psi|^2\sqrt{\det g}$, where $g$ is the metric tensor in the collective
space.

Table~\ref{tab:super} contains more detailed information on the calculated
properties of the lowest normally and super deformed states in the
\nuc{240}{Pu} nucleus. We show here also the average values $\langle
\beta\rangle$ and $\langle \gamma\rangle$.
\begin{table}[htb]
\tbl{Lowest normal and superdeformed states $J=0$ and 2 in
$^{240}$Pu (in MeV).\label{tab:super}}
{\begin{tabular}{cccrrrr@{\hspace{1cm}}r@{\hspace{0.3cm}}r}\toprule
  &J&  \#\tabmark{a} &  \multicolumn{1}{c}{$\langle\beta\rangle$} & 
\multicolumn{1}{c}{$\langle\gamma\rangle$} & \multicolumn{1}{c}{$E_{\rm th}$}
& \multicolumn{1}{l}{$E_{\rm exp}$}
& \multicolumn{1}{l}{($E-E_{0, \rm SD})_{\rm th}$}
& \multicolumn{1}{c}{($E-E_{0, \rm SD})_{\rm exp}$}\\
\colrule
ND & 0 &  0 &  0.299 &  8.09      & \\
&  2 & 0 &  0.299 &  8.07 &  0.046&  0.043\\
&  0 & 1 &  0.315 & 10.54 &  1.420 & 0.860\\
&  2 & 2 &  0.315 & 10.50 &  1.470 & 0.900\\
&  2 & 1 &  0.314 & 13.07 &  1.347 & 1.137\\
\\
SD&  0&  7 &  0.829 &  2.94& 4.321& 2.55 &        &\\
  &  2& 12 &  0.829 &  2.94& 4.342& & 0.021 \\
  &  0& 11 &  0.836 &  3.52& 5.472& &  1.151 & {}{0.77}\\
  &  2& 19 &  0.835 &  3.53& 5.494& &  1.173 \\
\botrule \end{tabular} }
\begin{tabfootnote}
\tabmark{a} \# denotes the number of a state for a given spin.
\end{tabfootnote} 
\end{table}

Obtained qualitative agreement with experiment looks rather encouraging.
Moreover, one must remember that we have not included in our calculation the
so called rotational correction often discussed in papers on fission
barriers.\cite{2004BO34,2007BA17} Its magnitude increases with a deformation
and in consequence superdeformed states can be lowered by 0.8-0.9 MeV.
However in our opinion a consistent introduction of this correction into our
formalism (keeping in view also allowed triaxial shapes of a nucleus) needs
more careful treatment.

\section{Conclusions} 
The ATDHFB approach offers a considerable extension of
the area of applicability of the self-consistent mean field theory, namely
on low energy collective states, for which there is often rich spectroscopic
data. Our study shows that starting with standard Skyrme interactions one
can obtain reasonable description of energies and B(E2)'s also for very
heavy nuclei. However some problems still remain open, e.g. questions
concerning the Thouless-Valatin corrections, pairing vibrations, rotational
and other so called zero point energy corrections.

\section*{Acknowledgments}

Present work was partially supported by the Academy of Finland and
Jyv\"askyl\"a University within the FiDiPro programme.

\end{document}